\documentclass[12pt]{article}

\usepackage[letterpaper,hmargin=1in,vmargin=1in]{geometry}

\usepackage{graphicx,epstopdf,amsmath,amsfonts,amssymb,cite}

\def\be{\begin{equation}}
\def\ee{\end{equation}}
\def\ba{\begin{eqnarray}}
\def\ea{\end{eqnarray}}
\def\ge{\mathrel{\raise.3ex\hbox{$>$\kern-.75em\lower1ex\hbox{$\sim$}}}}
\def\la{\mathrel{\raise.3ex\hbox{$<$\kern-.75em\lower1ex\hbox{$\sim$}}}}

\def\simgt{\mathrel{\raise.3ex\hbox{$>$\kern-.75em\lower1ex\hbox{$\sim$}}}}
\def\simlt{\mathrel{\raise.3ex\hbox{$<$\kern-.75em\lower1ex\hbox{$\sim$}}}}

\newcommand{\bi}[1]{\bibitem{#1}}
\newcommand{\fr}[2]{\frac{#1}{#2}}

\newcommand{\nc}{\newcommand}

\nc{\gone}{\bar g_{\pi NN}^{(1)}}
\nc{\gzero}{\bar g_{\pi NN}^{(0)}}
\nc{\al}{\alpha}
\nc{\ga}{\gamma}
\nc{\de}{\delta}
\nc{\ep}{\epsilon}
\nc{\ze}{\zeta}
\nc{\et}{\eta}
\nc{\ka}{\kappa}
\nc{\rh}{\rho}
\nc{\si}{\sigma}
\nc{\ta}{\tau}
\nc{\up}{\upsilon}
\nc{\ph}{\phi}
\nc{\ch}{\chi}
\nc{\ps}{\psi}
\nc{\om}{\omega}
\nc{\Ga}{\Gamma}
\nc{\De}{\Delta}
\nc{\La}{\Lambda}
\nc{\Si}{\Sigma}
\nc{\Up}{\Upsilon}
\nc{\Ph}{\Phi}
\nc{\Ps}{\Psi}
\nc{\Om}{\Omega}
\nc{\ptl}{\partial}
\nc{\del}{\nabla}
\nc{\ov}{\overline}
\nc{\newcaption}[1]{\centerline{\parbox{15cm}{\caption{#1}}}}
\nc{\us}{U(1)$_S$}

\def\beq{\begin{equation}}
\def\eeq{\end{equation}}
\def\bmat{\begin{displaymath}}
\def\emat{\end{displaymath}}
\def\bear{\begin{eqnarray}}
\def\eear{\end{eqnarray}}
\def\ba{\begin{eqnarray}}
\def\ea{\end{eqnarray}}
\def\bery{\begin{array}}
\def\ery{\end{array}}
\def\bit{\begin{itemize}}
\def\eit{\end{itemize}}
\def\ben{\begin{enumerate}}
\def\een{\end{enumerate}}
\def\btab{\begin{tabular}}
\def\etab{\end{tabular}}
\def\btbl{\begin{table}}
\def\etbl{\end{table}}
\def\bfig{\begin{figure}[htb]}
\def\efig{\end{figure}}
\def\bpic{\begin{picture}}
\def\epic{\end{picture}}

\def\ga{\mathrel{\raise.3ex\hbox{$>$\kern-.75em\lower1ex\hbox{$\sim$}}}}
\def\la{\mathrel{\raise.3ex\hbox{$<$\kern-.75em\lower1ex\hbox{$\sim$}}}}
\def\gappeq{\mathrel{\rlap {\raise.5ex\hbox{$>$}}
{\lower.5ex\hbox{$\sim$}}}}
\def\lappeq{\mathrel{\rlap{\raise.5ex\hbox{$<$}}
{\lower.5ex\hbox{$\sim$}}}}

\def\gyr{{\rm \, G\kern-0.125em yr}}
\def\mev{{\rm \, Me\kern-0.125em V}}
\def\gev{{\rm \, Ge\kern-0.125em V}}
\def\tev{{\rm \, Te\kern-0.125em V}}

\def\slash#1{\rlap{\hbox{$\mskip 1 mu /$}}#1}%

\begin{document}

\begin{titlepage}

\setcounter{page}{1}

\vspace*{0.2in}

\begin{center}

\hspace*{-0.6cm}\parbox{17.5cm}{\Large \bf \begin{center}

Multi-lepton Signatures of a Hidden Sector \\ in Rare \boldmath$B$ Decays

\end{center}}

\vspace*{0.5cm}
\normalsize

\vspace*{0.5cm}
\normalsize

{\bf Brian Batell$^{\,(a)}$, Maxim Pospelov$^{\,(a,b)}$, and Adam Ritz$^{\,(b)}$}

\smallskip
\medskip

$^{\,(a)}${\it Perimeter Institute for Theoretical Physics, Waterloo,
ON, N2J 2W9, Canada}

$^{\,(b)}${\it Department of Physics and Astronomy, University of Victoria, \\
     Victoria, BC, V8P 1A1 Canada}

\smallskip
\end{center}
\vskip0.2in

\centerline{\large\bf Abstract}

We explore the sensitivity of flavour changing $b\to s$ transitions 
to a (sub-)GeV hidden sector with generic couplings  to the
Standard Model through the Higgs, vector and axion  portals. 
The underlying two-body decays of $B$ mesons, $B\to X_s {\cal S}$ and 
$B^0 \to  {\cal SS}$, where ${\cal S}$ denotes a generic new GeV-scale particle, may significantly enhance the 
yield of monochromatic lepton pairs in the final state via prompt ${\cal S}\to l\bar l$ decays. 
Existing measurements of 
the charged lepton spectrum in neutral-current semileptonic $B$ decays
provide bounds on the parameters of the light sector that are significantly more stringent than the 
requirements of naturalness. New search modes, such as $B\to X_s +n(l\bar l)$ and 
$B^0 \to n(l\bar l)$ with $n\geq 2$, can provide additional sensitivity to scenarios in which
both the Higgs and vector portals are active, 
and are accessible to (super-)$B$ factories and hadron colliders. 

\vfil
    
\end{titlepage}

\subsection*{1. Introduction}

The study of $B$ mesons at the $B$-factories, BaBar \cite{babardet}  and Belle \cite{belledet}, and the Tevatron experiments \cite{cdf,d0}
has significantly advanced the precision with which various Standard Model (SM) 
parameters are known, and consequently has placed stringent constraints on models of new physics
affecting quark flavour \cite{Review}. The prevailing view is that such new physics must reside 
at or above the electroweak scale, manifesting at low energies in modifications to the Wilson coefficients of
effective flavour-changing operators that arise once the heavy degrees of freedom 
are integrated out. Experimental precision, and the ability to make accurate
SM predictions, are thus the controlling factors in probing weak-scale new physics through precision
flavour observables. 

While new states charged under the SM are generically required to be rather heavy, 
light (sub-)GeV mass states in a hidden sector, {\em neutral} under the SM gauge group, 
can peacefully co-exist with the 
SM, evading precision flavour and  electroweak constraints. Such hidden sectors may be weakly coupled
to the SM in various ways, and are often  best probed via experiments at the luminosity frontier. 
In particular, precision  studies of rare SM decays can provide impressive sensitivity to 
these sectors, opening the possibility for novel decay channels not encountered in the SM itself. Indeed, 
 over the years  there have been numerous searches for rare decays of  flavoured mesons to new light
 states (see {\em e.g.} \cite{old2} for a subset of theoretical ideas). As one notable motivation, these 
 hidden sector states can have a significant impact on Higgs decay channels, allowing
 for a SM-like Higgs with mass well below the conventional LEP bound \cite{nsh}.

In this paper, we revisit the sensitivity of rare flavour-changing decays 
from the generic standpoint of `portal' operators \cite{pw,portal}, which constitute 
a systematic way to parametrize the allowed couplings of  generic neutral states ${\cal S}$ in a hidden sector 
to the SM in  order of increasing canonical operator dimension. 
In particular, we will be interested in the following set of lowest-dimension portals:
\begin{eqnarray}
&&H^\dagger H (AS + \lambda S^2)~~~~~~~~~~~{\rm Higgs~portal~(dim=3,4),} \nonumber\\
&&\kappa F^Y_{\mu\nu}F'_{\mu\nu}~~~~~~~~~~~~~~~~~~~~~~{\rm Vector~portal~(dim=4),} 
\label{portals} \\ 
&&Y_N\bar L H N ~~~~~~~~~~~~~~~~~~~~~~{\rm Neutrino~portal~(dim =4)}, \nonumber \\
&&f_a^{-1}\bar \psi \gamma_\mu\gamma_5 \psi \partial_\mu a ~~~~~~~~~~~~~~ {\rm Axion~portal~(dim =5)}. \nonumber
\end{eqnarray}
Here $H$ is the SM Higgs doublet, $F_{\mu\nu}^Y$ is the hypercharge field strength, $L$ is the left-handed lepton doublet, and 
$\psi $ is a generic SM fermion, while ${\cal S} = S,~N,~ A_\mu'$ and $a$ denote the fields associated with new light states. 
The purpose of this study is to analyze the feasibility of searching for light states coupled to the SM via these 
portals in $B$ meson decays.\footnote{Renewed interest in the possibility of light hidden sector 
states coupled to the SM has emerged from attempts to link certain unexpected 
features in the multi-GeV scale cosmic electron and positron spectra to the 
annihilation of galactic dark matter into such light states \cite{pamela,astro}.}
Specifically, we will concentrate on the manifestations of Higgs, vector, and axion portals in $b\to s$ transitions with the
direct production of one or more exotic states. To be as conservative as possible, we shall not assume any direct flavour-violating 
operators, which in fact is automatic for the Higgs and vector portals, but requires an extra assumption for the axion portal. 
Using the resulting flavour-blind portal operators, we calculate the 
strength of the flavour-changing transitions induced by SM loops. 

A primary feature that we will exploit is that the scalar and axion (i.e. axial-vector) portals behave very differently to the conserved 
vector current portal once dressed by $W-(u,c,t)$ loop corrections. Schematically, this difference can be illustrated as follows:
\be
\bar t \gamma_\mu t \longrightarrow (G_F q^2) \times  \bar b_L \gamma_\mu s_L; 
~~~ \bar t \gamma_\mu \gamma_5t \longrightarrow (G_F m_t^2) \times \bar b_L \gamma_\mu s_L.
\label{enhancement}
\ee
While conservation of the vector current (such as the electric charge or baryon number) 
requires the dependence on $q^2 \la m_b^2$, the axial current is not conserved 
and the vertex correction is ${\cal O}(m_t^2/q^2)$--enhanced relative to the vector case. Within the SM, 
scalar or axial-vector currents are associated purely with couplings to the $Z$ boson and the SM Higgs, which cannot be produced 
in on-shell $B$ decays. Thus, having light states in the spectrum with (pseudo)scalar or axial-vector couplings can 
enhance the loop-induced two-body decays of the $b$ quark by many orders of magnitude. 
The enhancement of the loop-induced SM Higgs coupling has been known for some time \cite{Willey,Randall}.
More recently it has been exploited in the context of $B$ 
meson decays to a pair of 
light dark matter particles through the Higgs portal \cite{BJKP},  decays to a singlet scalar mixed with 
the Higgs \cite{wise}, and decays to a light pseudoscalar in the NMSSM \cite{nmssm}. 
Rare Kaon decays to metastable mediators were considered in \cite{PRV,Pospelov}.

We will analyze a number of semi-leptonic and fully leptonic $B$ decay modes opened up by portal couplings, which
can serve as a powerful probe of new light states. As often happens in models of this type with intermediate cascade 
decays, the increased multiplicity of final state leptons implies minimal additional suppression
 \cite{SZ,AW,Pospelov}, thus enhancing  signal over background. 
Specifically, we calculate $B\to K(K^{*})S\to K(K^{*})l\bar l$ and $B^0\to SS \to 2(l\bar l)$
in the minimal extension of the SM by one real scalar $S$, and $B\to K(K^{*})a\to K(K^{*})l\bar l$ in the axion portal model. We will
show that the constraints imposed by $B$-physics in the kinematically accessible range where the leptonic decays of $S$ and 
$a$ occur within the detector are easily the most stringent experimental limits. We also extend our analysis to include 
the vector portal, and in particular the natural combination of Higgs and vector portals, and calculate the
branching of $B\to VK(K^{*})$, $B^0 \to VV$ 
and $B\to h'h' $. The final state of two Higgs $h'$ bosons of the extra U(1) group may be dominated by eight leptons. 
The most important point of our analysis is to show that multilepton signatures of $B$ meson decays, like 
$B^0 \to \mu^-\mu^+\mu^-\mu^+$, can be explored using {\em existing} datasets collected at the $B$ factories
and the Tevatron, providing significant new probes of these models with exotic light neutral states. 

The remainder of this paper is structured as follows. In section 2, we analyze rare $B$ decays in 
the minimal extension of the SM by a singlet scalar interacting through the Higgs portal, as well as an extension 
with a pseudoscalar singlet coupled via the axion-portal. Section 3 considers rare $B$ decay modes 
proceeding via a combination of Higgs and vector portals, and we present our conclusions in section 4.

\subsection*{2. Rare $B$-decays through the Higgs and axion portals}

The extension of the SM by a singlet scalar has been considered on numerous occasions, e.g.
for cosmological applications as a minimal model of dark matter, with stability imposed by symmetry \cite{singletDM,BPV}, 
or its impact on electroweak baryogenesis or inflation \cite{cosmo}. Novel experimental signatures, including extra decay channels 
for the SM Higgs boson, were addressed in \cite{Krasnikov1,EZ,BPV,wise,wells,displaced,Barger,nsh}. 

A generic renormalizable scalar potential that includes $S$ self-interactions 
and couplings to the SM via the Higgs portal is given by,
\be
\label{starting}
V = \lambda_4 S^4 + \lambda_3 S^3 + m_0^2S^2 + (AS+\lambda S^2)(H^\dagger H).
\ee
Since we are interested only in the low-energy limit of the theory relevant for $B$ decays, we will assume stability of the potential in the
$S$-direction and integrate out the Higgs boson to obtain an 
effective Lagrangian for $S$ (enforcing $\langle S \rangle =0$ by an appropriate shift of the field),
\be
\label{S}
{\cal L}_S = \fr12(\partial_\mu S)^2 - \fr12m_S^2S^2 - 
\left(\fr{\theta S}{v} + \fr{\lambda S^2}{m_h^2}\right){\cal L}_m - \fr{A'}{6}S^3 + \cdots.
\ee
The quantity ${\cal L}_m$ comprises the SM mass terms from electroweak symmetry breaking ({\em i.e.} ${\cal L}_m=m_l \bar ll+\cdots$), and 
the physical mass $m_S$, mixing angle $\theta$, and self-interaction parameter $A'$ are related to the parameters in 
(\ref{starting}). 
 The precise nature of these relations ($\theta \simeq A v/ m_h^2$ etc.) will not be critical to our analysis. However, the 
technical naturalness of the model (\ref{S}) is a valuable criterion to use in setting the characteristic values of $\theta$ and $A'$. 
In order to shelter a relatively light scalar from large mass corrections induced by electroweak symmetry breaking, we take
\ba
\theta &\la& \fr{m_S}{m_h} \sim {\cal O}(10^{-2}) \times \left(\fr{m_S}{1~\rm GeV}\right), \nonumber \\
A' &\la& (16 \pi^2 m_S^2)^{1/2} \sim {\cal O}(10~{\rm GeV}) \times \left(\frac{m_S}{1~{\rm GeV}}\right).
\label{natural}
\ea
The latter relation follows from the $SS$ loop correction to the mass of the scalar. A larger angle $\theta$ and 
self-interaction parameter $A'$
would require additional tuned cancellations between different contributions to $m_S$. 
The possibility of a stronger coupling to the Higgs portal,
while keeping $S$ light and avoiding the naturalness constraints, arises in the large 
$\tan\beta$ two-Higgs doublet extension of the SM \cite{BJKP} and thus also in the MSSM.

We will also explore the axion portal, which avoids corrections to the (sub-)GeV mass of the pseudoscalar 
via the dimension-five axial-vector couplings of the form,
\be
\label{La}
{\cal L}_a~~ =  \sum_{{\rm SM-}\psi}\fr{\partial_\mu a}{f_\psi}\bar\psi \gamma_\mu \gamma^5 \psi.
\ee
Furthermore, for simplicity, we will neglect the effects of the self-interaction of $a$, as well as couplings
to gauge bosons, and assume universal couplings 
of the pseudoscalar to leptons $f_l$  and quarks, $f_q$. This automatically protects (\ref{La}) 
from tree-level flavour changing neutral currents (FCNCs). While a UV completion is required 
for (\ref{La}), we note that in two-Higgs doublet extensions of the SM there also exists the possibility of a 
renormalizable pseudoscalar portal, e.g. $i a H_1H_2$, which leads to the mixing of $a$ with 
the pseudoscalar Higgs boson $A$.

For both the Higgs and axion portals, on integrating out the $W$-top  loop, we obtain the well-known effective $b-s-h$ and $b-s-a$ vertices,
\be
{\cal L}_{bs} = \frac{3\sqrt{2} G_F m_t^2 V_{ts}^* V_{tb}}{16 \pi^2 } \times m_b \bar s_L b_R \times \left ( \fr{\theta S}{v}  - 
i\fr{2}{3}\fr{a}{f_q}\ln\left(\La_{\rm UV}^2/m_t^2\right)    \right )
+({\rm h.c.})
\label{Leff}
\ee

For the scalar $S$, the Wilson coefficient in (\ref{Leff}) is one-loop exact in the limit $m_b^2/M_W^2\to 0$, 
while for the pseudoscalar\footnote{We thank the authors of Ref. \cite{Jesse} for pointing out the presence of a logarithmic UV divergence in this calculation.} we retain only the leading log-divergent term proportional to $m_t^2/m_W^2$ and for consistency assume at least a small hierarchy between the weak scale and the UV cutoff, $\ln{\La_{\rm UV}/m_t} \sim 1$.
We have integrated by parts and used the equations of motion for the quark fields in the limit $m_s=0$ 
to remove the derivative from the axion field in the interaction (\ref{Leff}). 

The Lagrangian (\ref{Leff}) immediately leads to the inclusive $b$ quark decay width to $S$ and $a$, but we are more interested in 
$K$ and $K^*$ final states. The QCD matrix elements involved in $B_{d(u)}$ to $K(K^*)$ transitions have been calculated using 
light-cone QCD sum rules \cite{QCDsr1,QCDsr2},
and after a fairly standard calculation, we obtain 
the following results as functions of $m_S$ and $m_a$:
\begin{eqnarray}
{\rm Br}_{B\to K S} & \simeq & 4 \times 10^{-7} \times \left( \fr{\theta}{10^{-3}} \right)^2 {\cal F}^2_K(m_S) \lambda_{KS}^{1/2} \nonumber\\
\label{results}
{\rm Br}_{B\to K^* S} &\simeq& 5 \times 10^{-7} \times \left( \fr{\theta}{10^{-3}} \right)^2 {\cal F}^2_{K^*}(m_S) \lambda_{K^*S}^{3/2}\\
{\rm Br}_{B\to K a}  &\simeq& 5 \times 10^{-6} \times\left( \fr{100~ {\rm TeV}}{f_q} \ln\left(\frac{\La_{\rm UV}}{m_t}\right)\right)^2 {\cal F}^2_K(m_a) \lambda_{Ka}^{1/2}\nonumber\\
{\rm Br}_{B\to K^* a} &\simeq &6 \times 10^{-6} \times \left( \fr{100~ {\rm TeV}}{f_q} \ln\left(\frac{ \La_{\rm UV} }{m_t}\right)\right)^2 {\cal F}^2_{K^*}(m_a) \lambda_{K^*a}^{3/2}.\nonumber
\end{eqnarray}
The dependence on the unknown mass parameters resides in the phase space factors, 
$\lambda_{ij} = (1-m_B^{-2}(m_i+m_j)^2)(1-m_B^{-2}(m_i-m_j)^2)$, and the form factors
which we have normalized to their values at zero momentum transfer \cite{QCDsr2},
\ba
{\cal F}_K(m) &=&  \fr{1}{1-m^2/(38~{\rm GeV}^2)}, \nonumber \\
{\cal F}_{K^*}(m) &=& \fr{3.65}{1-m^2/(28~{\rm GeV}^2)}
-\fr{2.65}{1-m^2/(37~{\rm GeV}^2)}.
\ea
The values of the form factors at $q^2=0$ used in our calculations are $f_0(0) = 0.33$ and $A_0(0) = 0.37$ \cite{QCDsr2}.
The uncertainty in the form factors is the main source of error for (\ref{results}), argued to be
at the ${\cal O}(10-15\%)$ level \cite{QCDsr1,QCDsr2}. 

The results in (\ref{results}), combined with the subsequent decay of $a$ or $S$ to dilepton pairs close to the interaction 
point presents an intriguing signal: a monoenergetic lepton pair in association with $K$ or $K^*$. The branching ratios 
${\rm Br}_{B\rightarrow K\mu\bar\mu} =  4.2_{-0.8}^{+0.9}  \times 10^{-7}$ 
and ${\rm Br}_{B\rightarrow K^*\mu\bar\mu} = 1.03_{- 0.23}^{+ 0.26}  \times 10^{-6}$ have been measured 
\cite{PDG,Babar,Belle} with several hundred decays containing lepton pairs distributed over the entire
available $q^2$ range, while a monoenergetic lepton pair can be 
efficiently probed at $B$-factories with ${\cal O}(10^{-8})$ sensitivity \cite{talk_DF}. The hadronic decays of $S$ and $a$ 
as well as missing energy signatures from decays outside the detector can also be probed, albeit with lesser sensitivity.

With $S$ and $a$ in the intermediate state, the decay widths and branching ratios to leptons are sensitive
functions of mass. We follow the standard prescriptions for calculating the total widths of $a$ and $S$
\cite{Voloshin,HHguide,TW}, and the results can be summarized as follows. When only decays to leptons are kinematically allowed, 
the leptonic branching is necessarily close to unity,\footnote{Within the axion portal scenario, for certain parameter choices the 
decay to $\gamma\gamma$ can be significant and may also be a good search mode.} while the partial decay 
width to a lepton pair is given by,
\be
\Gamma_{S\to l\bar l} = \fr{\theta^2m_l^2m_S}{8\pi v^2}\left(1-\fr{4 m_l^2}{m_S^2}\right)^{3/2}  ,~~~ 
\Gamma_{a \to l\bar l} =  \fr{m_l^2m_a}{8\pi f_l^2}\left(1-\fr{4 m_l^2}{m_a^2}\right)^{1/2},
\ee
and is very sensitive to whether the dimuon channel is open. For example, for a $250$ MeV mass scalar with mixing angle $10^{-3}$ 
the lifetime is $c\tau = 2.7$ cm, and considering a Lorentz boost of $\gamma \sim m_B/(2m_S) \sim 10$, this would correspond to a 
significantly displaced vertex.

For higher mass scalars, the decay length shrinks while the leptonic branching gets suppressed, 
especially near the $f_0$ $0^+$ resonance \cite{TW}. In the region near the resonance, we base our estimate of the branching on 
a coupled-channel analysis in the framework of chiral perturbation theory, while above the resonance, we use 
perturbative QCD \cite{Voloshin,TW}:
\begin{eqnarray}
{\rm Br}_{S\to \mu\bar \mu} &\sim & \fr{  m_\mu^2 \beta_\mu^3 }
{m_\mu^2 \beta_\mu^3 + | F_\pi/2 m_S|^2 \beta_\pi +| F_K/2 m_S|^2 \beta_K}
~~\quad {\rm for}~~ m_S \la 1.5 \, {\rm GeV},\nonumber\\
{\rm Br}_{S\to \mu\bar \mu} &\sim & \fr{m_\mu^2}
{m_\mu^2 + 3m_s^2 + m_S^2 (\alpha_s/\pi)^2 (N_f^2/9) +\cdots }
~~~~~ \quad{\rm for} ~~ m_S >1.5 \, {\rm GeV},  
\end{eqnarray}
where $\beta_i=(1-4m_i^2/m_S^2)^{1/2}$, $F_i$ are the form-factors defined in Ref.~\cite{TW}, 
$N_f$ is the number of heavy quarks, {\it i.e.} three below the charm threshold,
and the ellipsis in the second line stands for charm and $\tau$ contributions once the corresponding thresholds are open.
We note that there is at least a $100\%$ uncertainty in this formula above 1 GeV \cite{Voloshin}.

For the pseudoscalar case, the hadronic width is suppressed by  three
pion phase space. In order to estimate the scaling of the branching ratio with $f_l$ and $f_q$ we
assume that the decay to hadrons occurs via mixing with the $\eta$ and $\eta'$ resonances.
Taking a representative value of $m_a = 800 $ MeV, 
the mixing with $\eta'$ is given by
$\theta_{a\eta'} \sim (f_{\eta'}/f_q)\times
\sqrt{3}m_a^2/(m_{\eta'}^2-m_a^2)$,
and the hadronic width is approximately
$\Gamma_{\rm had} \sim \theta_{a\eta'}^2\Gamma_{\eta'}$.
Using these results, we obtain the following scaling of the 
leptonic branching fraction:
\begin{eqnarray}
{\rm Br}_{a\to \mu\bar \mu} \sim \fr{1}{1+ 0.3 (f_l/f_q)^2}.
\end{eqnarray}
It is apparent that the resonant enhancement of the
hadronic width can significantly exceed the
naive three-pion continuum result.

With these estimates in hand, we can predict the observable signal at (super-)$B$ factories. Having a typical 
detector design in mind, we require $S$ or $a$ to decay 
within a transverse distance $l_{\rm min}=25$ cm of the beam pipe, and assume $\sim 90$\% angular acceptance. 
In practise this amounts to calculating the following angular integral multiplying Eqs. (\ref{results}):
\be
{\rm Br}_{S(a)\to \mu\bar\mu}\int_{\theta_{\rm min}}^{\pi-\theta_{\rm min}}\fr{\sin\theta d\theta}{2}
\left( 1  - \exp\left[ -\fr{l_{\rm min}\Gamma_{S(a)}}{\gamma_{S(a)}\sin\theta }  \right]    \right). 
\ee
In the limit of a short decay length, the integral is trivially 
${\rm Br}_{S(a)\to \mu\bar\mu}\cos\theta_{\rm min}$, and in the opposite limit of a very long 
decay length it is $ (\Gamma_{S(a)\to \mu\bar \mu} \,l_{\rm min} ) \times (\pi/2-\theta_{\rm min})\gamma_{S(a)}^{-1} $.

\begin{figure}
\centerline{
\includegraphics[width=.5\textwidth]{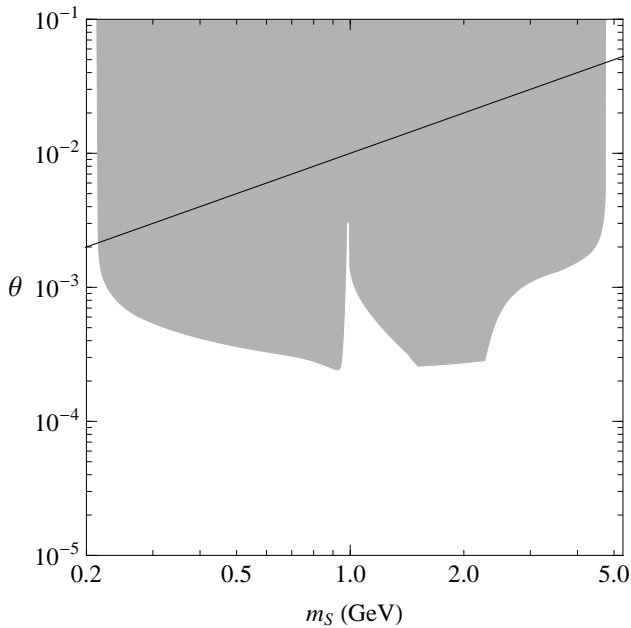}}
\caption{\footnotesize Sensitivity of the BaBar/Belle dataset to combined $B\rightarrow KS$ and 
$B\rightarrow K^*S$ decays in the dimuon channel. The region below the dashed line is technically 
natural as discussed in Eq.~(\ref{natural}).}
\label{figBKS1}
\end{figure}

\begin{figure}
\centerline{
\includegraphics[width=.5\textwidth]{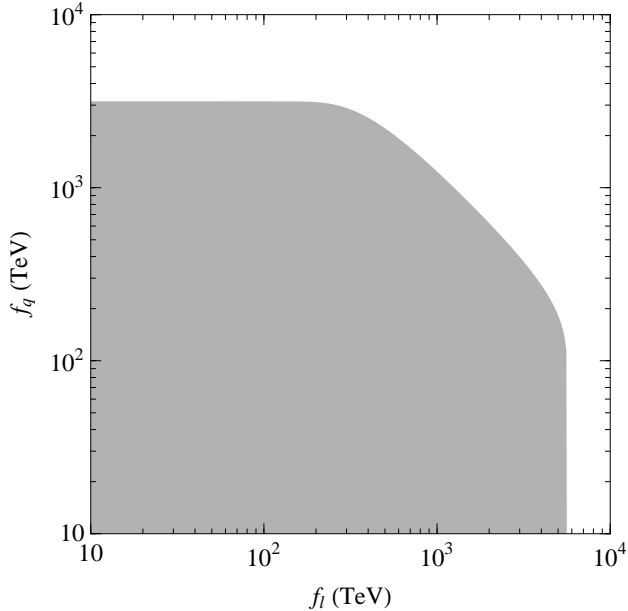}}
\caption{\footnotesize Sensitivity of the BaBar/Belle dataset to combined 
$B\rightarrow Ka$ and $B\rightarrow K^*a$ decays in the dimuon channel. We have set $\ln{\La_{\rm UV}/m_t} \sim 1$.}
\label{figBKS2}
\end{figure}

Given that the combined BaBar/Belle dataset provides sensitivity to the 
 $K\mu\bar \mu $ and $K^*\mu\bar \mu$ branching
with a mono-energetic 
muon pair at the level of ${\cal O}(10^{-8})$, the significant parameter space reach that ensues for
the two models is shown in Figs.~\ref{figBKS1} and \ref{figBKS2}.
For the scalar singlet Higgs portal, Fig.~\ref{figBKS1} illustrates that the $B$-factories can probe deep within the 
technically natural region of the $\theta-m_S$ parameter plane (see Eq.~(\ref{natural})), with sensitivity to mixing 
angles in the $10^{-4}-10^{-3}$ range. For light scalars with masses below the $2\pi$ threshold we see that, 
although the branching to dimuons approaches 100\%, the sensitivity is diminished as the $S$ particle is very narrow and 
long-lived and thus able to escape the detector. 
We also observe that the sensitivity is weakened near the $f_0$ resonance, and for heavy scalars, as in these regions the 
branching to muons is small. For the axion portal, 
we present in Fig.~\ref{figBKS2} the $f_q-f_l$ sensitivity for an 800 MeV pseudoscalar, indicating that
the sensitivity to the axion couplings reaches $f_{q,l} \sim 10^3$ TeV. Qualitatively, we see that 
when $f_q$ is large, sensitivity is lost as the branching of $B$ mesons to pseudoscalars 
is small, while for large $f_l$ sensitivity is lost as the decays of $a$ are primarily hadronic.
Nonetheless, we note that the sensitivity to axion couplings obtained here appears significantly stronger than 
that of Ref.~\cite{Jesse}. We believe that much of this numerical discrepancy can be attributed to the 
difference in experimental sensitivity to the branching fraction used in the two analyses. In addition, we assume at least a small hierarchy exists between the weak scale and the
UV cutoff, whereas Ref.~\cite{Jesse} considers the UV-complete two-Higgs-doublet model,
in which - without this hierarchy - the top-$W$ loop has
an additional suppression compared to Eq.~(\ref{Leff}).  
Finally, it is also important to emphasize the complementarity of  constraints from rare $K$ and $B$ decays. 
For a weakly interacting (pseudo)scalar particle with a mass below the dimuon threshold
and a long lifetime, the $K\to \pi + \slash{\!E}$ decay  (e.g. $K\to \pi \nu \bar{\nu}$) is the most efficient probe \cite{PRV}. 
On the other hand, a semi-leptonic signature of $S$ or $a$ is more efficiently probed via $B$ decays, since the CKM suppression 
from the top-loop is less severe.

There are several other interesting signatures for the Higgs portal scenario in Eq.~(\ref{Leff}). 
Consider the decay of  $B^0$ mesons to a pair of scalars. Assuming for simplicity that the $A'$ trilinear vertex 
dominates, we obtain the following estimate for the branching to an $SS$ pair:
\begin{eqnarray}
\label{2S}
{\rm Br}_{B_s \to SS} \simeq
4\times 10^{-3} \times \theta^2  \left(\fr{A'}{m_B}\right)^2  \fr{\lambda_{SS}^{1/2}}{(1-m_S^2/m_B^2)^2}.
\end{eqnarray}
The suppression of the $2S$ final state relative to $KS$ is due to the fact that the decay amplitude for $2S$ 
is proportional to the decay constant $f_B \simeq 200$ MeV, while the $KS$ decay amplitude, in the same 
units, is controlled by $f_0 m_B \sim $ 2 GeV. For  $B_d$ decays there is of course an extra CKM 
suppression by $|V_{td}/V_{ts}|^2$ relative to (\ref{2S}). Nonetheless, the overall rate to muons for 300 MeV scalars can reach 
${\rm Br}_{B_S\to 4\mu} \sim 10^{-8}$ with a moderate fine-tuning of couplings to allow for a larger $A'$.

Returning to the decays mediated by $\lambda SSH^\dagger H$, we note that only in the limit 
$\lambda \ga 10^{-2}$ is the branching for $B_s\to 4 \mu$  above the $10^{-8}$ level. Such values of 
$\lambda$ are difficult to reconcile with the large additive renormalization of $m_S^2$ by $\lambda v^2$, 
which would require fine tuning at the level of 1 part in $10^3$ for a 1 GeV scalar. Such a fine tuning can be avoided 
in the two-Higgs doublet model with a portal $\lambda_{H_1}H_1^\dagger H_1 S^2$, where 
$\lambda_{H_1}$ can naturally be $ {\cal O}(1)$ if $\tan \beta$ is maximal, 
$\tan \beta = \langle H_2 \rangle / \langle H_1 \rangle \sim 50$. Taking the results of the
$b-s-S^2$ transition calculated in \cite{BJKP}, with a charged Higgs mass $m_{H^+} = 300$ GeV, 
we obtain the following estimate for the rate of the $B_s \to 4 \mu $ transition:
\be
\label{4mu}
{\rm Br}_{B_s\to 2S\to 4\mu} \simeq 2\times 10^{-7} \times \lambda_{H_1}^2  \lambda_{SS}^{1/2} \times  {\rm Br}_{S\to 2\mu}^2.
\ee
Assuming a similar sensitivity to the four-muon channel as for $\mu\bar \mu$ at CDF \cite{CDF}, we conclude that the Tevatron experiments
can probe $\lambda_{H_1}^2 \times {\rm Br}_{S\to 2\mu}^2$ at the ${\cal O}(0.1)$ level.  A tension in the parameters arises 
if (\ref{4mu}) is to be maximized: larger values of $\lambda_{H_1}$ imply larger values of $m_S$ where 
${\rm Br}_{S\to 2\mu}$ diminishes. If ${\rm Br}_{S \to 2\mu} \ll 1$, searches for $l\bar l \pi^+\pi^-$ and 
$l \bar l K^+K^-$ final states with two hadrons reconstructing  the same invariant mass 
might be more advantageous than the search for fully leptonic decays of both $S$ scalars.

\subsection*{4. Rare $B$-decays through the \us\ sector}

In this section we will discuss $B$-decays via the combined Higgs and vector portals,
\be
\label{combo}
{\cal L}_{\rm Higgs+Vector} =- \lambda (H^\dagger H)(H^{'\dagger} H') - \fr{\kappa}{2}F_{\mu\nu}F'_{\mu\nu},
\ee
where $H'$ is a new scalar field charged under an additional \us\, gauge group, while the SM is \us-neutral. 
The vector portal in (\ref{combo}) 
is the minimal possibility \cite{Holdom} although other options that involve the gauging of anomaly-free SM quantum  
numbers are also plausible \cite{otherU1}. The gauging of the scalar coupled via the Higgs portal has two important 
consequences. First, as has been emphasized in many papers (see, e.g. \cite{Wells,Pospelov,BPR}), the yield of leptons 
in the final state can be enhanced, as the decay of the physical excitation $h'$ may proceed via the intermediate 
vector states of \us\, which in turn cascade to leptons:
\be
h'\to VV\to l\bar l l  \bar l.
\label{hVV}
\ee 
The vectors decay with equal probability to different (charged) lepton species, so that the decay to electrons is no longer suppressed. 
The decay chain (\ref{hVV}) is efficient if $m_{h'} >2 m_V$, and the relative branching of $V$ to leptons 
for the minimal portal is regulated by the well-measured process $\gamma^*\to{\rm hadrons}$ \cite{BPR}
characterized by the $R(s)$ ratio. A second important consequence is that the decay chain (\ref{hVV}) is likely to be very prompt, 
occurring within the detector even for very small values of $\kappa$.

We first address $B\to KV$ decays within the pure vector portal model. In this case, on account 
of (\ref{enhancement}), there is no particular enhancement. Calculation of the decay width
involves the familiar $Z$ and $\gamma$ penguins, with the vector particle attached via
kinetic mixing.  The result turns out to be very small, and for $m_V\sim 1$ GeV we find,
\be
{\rm Br}_{B\to KV} \sim 6\times 10^{-7} \kappa^2.
\ee
This channel appears to be less sensitive to the kinetic mixing parameter $\kappa$ than existing 
limits from other low-energy precision
experiments \cite{Pospelov,BPR,BPR2}\footnote{Further model-dependent sensitivity to the kinetic mixing parameter 
may be obtained with cosmic- and gamma-ray experiments and neutrino telescopes \cite{LLP}.}.

The next process we consider is $B\to K(K^*)h'\to K(K^*) VV \to K(K^*)l\bar l l  \bar l$. Utilizing the
results (\ref{results}), we obtain
\be
\label{K4l}
{\rm Br}_{B\to K(K^*)l\bar l l\bar l} \simeq 0.5\times \left( \fr{\lambda v'v}{m_h^2} \right)^2 \fr{1}{(1+R(m_V)/2)^2 },
\ee
having assumed that $\Gamma_{V\to e\bar e}  = \Gamma_{V\to \mu\bar\mu}$. From (\ref{K4l}) one can infer rather
strong ${\cal O}(10^{-4})$ sensitivity to the mixing parameter $\lambda vv'm_h^{-2}$. However, it is important to
bear in mind that the naturalness limits on $\lambda$ are also quite strong, $\lambda v^2 \la {\cal O}(m_{h'}^2)$,
and therefore (\ref{K4l}) is not probing the natural strength of the Higgs portal. 

A particularly interesting aspect of the combined Higgs and vector portals is that the decay $ B^0\to VV$
can proceed through an off-shell $h-h'$ propagator. At first, it may appear that this process is insignificant, 
as both $h-h'$ mixing and the $h'-V-V$ vertex are proportional to $v'$, naively suggesting strong suppression 
for a light vector. However, it turns out that the longitudinal vector modes in the final state cancel 
this $v'$-dependence so that the result remains finite in the 
$m_V \to 0$ limit,
\be
{\rm Br}_{B_s\to VV} = 4\times 10^{-5} \times \lambda^2  \lambda_{VV}^{1/2} \times
\fr{1-4m_V^2/m_B^2+12m_V^4/m_B^4}{(1- m_{h'}^2/m_B^2)^2},
\ee
where we have taken $m_h =115$ GeV. This decay leads to four leptons in the final state, 
and there is a possible enhancement of the rate 
for $m_{h'}$ close to $m_B$. 

Finally, the cascade decay $B\to 2h'\to 4V \to 4( l  \bar l)$ leads to eight leptons in the final 
state. The rate for this process may be enhanced in the two-Higgs doublet model, 
and reach ${\cal O}(10^{-7})\times \lambda_{H_1}^2$. Therefore probes of this signature at 
a level better than 1 part in $10^7$ at the Tevatron are well-motivated.

\subsection*{4. Conclusions}
We have shown that rare decays of $B$-mesons to semileptonic or fully leptonic final states can, via the
$B$-factory datasets, be a  sensitive probe for new light states coupled through the Higgs and axion portals. 
The results of sections 2  indicate that existing data allows for a probe of  neutral scalars coupled through the Higgs portal
down to mixing angles as small as $10^{-3}-10^{-4}$. In addition, the axion portal coupling to the top quark can be 
tested at an impressive level of sensitivity, $f_q \sim 10^3$ TeV. 

We have also shown that a combination of vector and Higgs portals, {\em e.g.} gauging of the 
scalar field coupled to $H^\dagger H$, can enhance sensitivity through the 
multilepton decays of the scalars. Among the novel signatures that we believe can be efficiently probed at both 
(super-)B factories and hadron colliders are the $K(K^*)+2(l\bar l)$, $2(l\bar l)$ and $4(l \bar l)$ final states. 
As far as we are aware, these final states have not been explored to date and thus represent a new
opportunity to access  light  mediators. 

Finally, we should mention that while we have focused on $B$-decays, and similar studies in the kaon sector have a long history, 
further sensitivity to these portal couplings may arise in the charm sector, via $D$-decays.

\subsubsection*{Acknowledgements}

We thank R. Kowalewski, Y. Kwon and M. Trott  for helpful discussions, and especially M. Freytsis, Z. Ligeti and J. Thaler for
emphasizing the UV-sensitivity of the $b-s-a$ vertex in Eq.~(7). We also thank 
the SLAC theory group for organizing the stimulating `Dark Forces workshop' in September 2009.
The work of A.R. and M.P. is supported in part by NSERC, Canada, and research at the Perimeter Institute
is supported in part by the Government of Canada through NSERC and by the Province of Ontario through MEDT.

\end{document}